\documentstyle[preprint,prb,aps]{revtex}
\input{epsf}
\tightenlines
\begin{document}
%%%%%%%%%1%%%%%%%%%2%%%%%%%%%3%%%%%%%%%4%%%%%%%%%5%%%%%%%%%6%%%%%%%%%7%%
%% TITLE
\draft
\title{Implementation of an all-electron  GW Approximation using the Projector 
Augmented Wave method: II. Application to the optical properties of semiconductors} 
\author{B.~Arnaud and  M.~Alouani}
%\author{B.~Arnaud, M.~Alouani, P.~Bl\"ochl}
\address{Institut de Physique et de Chimie des Mat\'eriaux de
Strasbourg (IPCMS), 23 rue du Loess,67037
Strasbourg Cedex, France}
\date{\today}
\maketitle

\begin{abstract}
We used our previously  implemented  GW approximation (GWA) based on  
the all-electron full-potential projector augmented wave (PAW) method to 
study the   optical properties of small, medium and large-band-gap 
semiconductors: Si, GaAs, AlAs, InP, Mg$_2$Si, C, and LiCl. The aim being to 
study the size of  both local-field (LF)  and the 
quasi-particle (QP) corrections   to  the calculated dielectric 
function obtained using the local density approximation
(LDA). We found that while the QP corrections tend to align 
the calculated structures in the optical spectra with their experimental 
counterparts, the LF effects  don't change these peak positions but
systematically reduce the intensities of the so called 
$E_1$ and $E_2$ structures in all the optical spectra. The reduction of the 
intensity of the $E_1$ peak worsen the agreement with experiment while that 
of $E_2$ improves it.  
We then show that the local-field correction improves considerably 
the calculated static dielectric constants of all studied semiconductors.
 Because the static dielectric constant is a ground state property,
the remaining discrepancy with experiment should be attributed to the 
the LDA itself. On the other hand, as expected, the calculation of the static
dielectric constant using the GW quasiparticle energies and including
the LF effects is underestimated 
for all the semiconductors. The  excitonic effects  should then 
correct for this discrepancy with experiment. 

\end{abstract}

%===================================================================
\section{Introduction}
%===================================================================
In a preceding paper\cite{brice} (referred to as I), we have implemented
the so called GWA\cite{Hedin1,Hedin2} using the all-electron projected 
augmented wave  and 
applied it successfully to compute the electronic structure of some small,
medium, and large-band-gap semiconductors: Si, GaAs, AlAs, InP, Mg$_2$Si, C,
and LiCl. We have then shown that the GWA accounts for most of the discrepancy
between the LDA and experiment regarding the energy position  of the  
conduction states. The remaining discrepancy with experiment is believed
to be due either to the way the decoupling of the core and valence electrons
is performed  or to the
non-selfconsistent implementation of our all-electron GW method.

 It is then clear that  the  GWA is  a 
useful method for calculating the quasiparticle properties of materials,
\cite{Hybertsen1,Hybertsen2,Godby1,Godby2,Godby3,vonderlinden,Horsch,ummels,bobbert,Hamada,Hott,arya,shirleylouie,shirleyzhu,review_gw1,review_gw2,aulbur_thesis,northruphl,saito,zhu,northrup,zhangclth,zhanghclt,shirleymartin,charlesworth}
and many applications using this method are now available. 
In particular, it has been successfully used to obtain a variety of physical 
properties, ranging form the band-width narrowing in alkali-metals, and their 
clusters,\cite{northruphl,saito}
 to the understanding of the surface reconstruction of semiconductors
\cite{zhu,northrup},  or the orientational disorder and the photoemission 
spectra of solid C$_{60}$.\cite{shirleylouie}

Nevertheless, one of the most  interesting quasiparticle application  
is the understanding of the  many-body effects on the 
single particle excitation's spectra, such as the optical spectra of 
semiconductors,  and it is still  relatively less explored. 
In this regards, The recent ``ab-initio'' inclusion of 
local-field and excitonic effects in 
the calculation of some semiconductor dielectric functions
is a great  achievement. \cite{albrecht,benedict,rohlfing1} 
There is, however, no systematic study of the trend  of
the size of these effects with respect to  several types of semiconductors. 
It is therefore important  to study the  
trend of  the  local-field contribution  to the intensities of 
the structures observed in the optical spectra of various types of 
semiconductors.

To date all  local-field calculations are based on pseudopotential methods.
In particular, Louie, Chelikowsky, and Cohen used the empirical 
pseudo-potential (EPP) method \cite{lcc} to computed the optical
properties of Si.  Recently Albrecht {\it et al.} used the ab-initio PP 
for    Si\cite{albrecht}, 
and Gavrilenko and Bechstedt for SiC, Si, and C\cite{gavrilenko}.
However,  these PP calculations do not agree with the  EPP\cite{lcc}, 
and do not provide  any definite  trend for the size of local-field 
contribution as a function of the semiconductor's type.

In this paper, we   use our newly developed  all-electron GWA 
based on  the  projector-augmented wave (PAW) method\cite{Blochl}
 to compute the local-field effect on various
types of semiconductors. The dynamical dielectric function is computed  
in the random-phase-approximation (RPA) with and without local-field effects using
both the LDA and quasiparticle energies calculated within our GWA method.  
The aim being a systematic study of the local
field effects in all these semiconductors in order to understand its 
contribution to the macroscopic dynamical dielectric function, to the
electron loss energy spectrum,  and to the 
static dielectric constant.  The latter being a ground state property, and
should, in principle, be well described by an LDA calculation including the
local-field effects. 

Our paper is organized as follows: In the second section  we introduce briefly 
our method of calculation. 
In the third section we apply it to determine the optical properties 
of two distinct semiconductor groups: 
some small and medium band gap semiconductors: Si, GaAs, AlAs, InP,
and Mg$_2$Si, and some large band gap semiconductors or insulators: C and LiCl. 
We then compare our results with available  calculations and experiments.
We will also  discuss the computed  values of the static dielectric constant 
using the  LDA with and without  local-field effects and the corresponding 
quasiparticle results.  This will lead us to  discuss the
  excitonic effect and its contribution to  the calculated quasiparticle
static dielectric constants.

\section{Method of calculation}
\subsection{Quasiparticles within the  GW approximation}
In this paper,  we extend 
the projector-augmented wave method(PAW)\cite{Blochl} 
for solve the Kohn-Sham equations to the  
determination of the optical properties of semiconductors. 
To make a quantitative comparison to
experimental results  we correct the LDA eigenvalues using the
quasiparticle energies. As described in our previous paper\cite{brice} 
we can find the excitation energies of the system by solving a 
quasiparticle equation
instead of locating the poles of the Green's function. 
The quasiparticle energies $E_n({\bf k})$ are determined from the electron
selfenergy operator $ \Sigma({\bf r},{\bf r}^{\prime},E_n({\bf k}))$:

\begin{equation}
(T+V_{ext}+V_{h})\psi_{{\bf k}n}({\bf r}) + \int d^3r^{\prime}
 \Sigma({\bf r},{\bf r}^{\prime},E_n({\bf k}))\psi_{{\bf k}n}({\bf
r}^{\prime})
=
 E_n({\bf k})\psi_{{\bf k}n}({\bf r})
\end{equation}
Here, T is the kinetic energy operator, $V_{ext}$ is the external
(ionic) potential, $V_{h}$
is the Hartree potential due to the average Coulomb repulsion of the
electrons. 
In the GWA, $\Sigma$ is  approximated
by a convolution with respect to the frequency variable of the Green's function, $G$,
with the screened interaction $W$ calculated within the RPA.

\subsection{Dielectric function and local field effect}

\subsubsection{Inclusion of local field effects at the RPA level}

In a crystal, which possesses lattice translation symmetry, a small electric 
perturbation $E_0({\bf q}+{\bf G},\omega)$ of wave vector 
${\bf q}+{\bf G}$ and frequency $\omega$  produces  responses 
$E({\bf q}+{\bf G}^{\prime},\omega)$ of wave vectors 
${\bf q}+{\bf G}^{\prime}$.  The  ${\bf G}$ and ${\bf G}^{\prime}$ being   
reciprocal lattice vectors. Thus, the dielectric matrix describing 
these  responses, is of the form 
$\epsilon_{{\bf G}^{\prime},{\bf G}}({\bf q},\omega)$ and it  can be written
as: 
\begin{equation}
E({\bf q}+{\bf G}^{\prime},\omega)=\sum_{{\bf G}}
 \epsilon^{-1}_{{\bf G}^{\prime},{\bf G}}({\bf q},\omega) E_0({\bf q}+{\bf G},\omega)
\end{equation}
An external macroscopic electric field can be viewed as a perturbation of 
vanishingly small
wave vector ${\bf q}$ and, therefore, the screening of the external macroscopic field
is given by the matrix element $\epsilon^{-1}_{{\bf 0},{\bf 0}}({\bf q},\omega)$ of
the inverse dielectric matrix. In insulating crystals, this results in a formula
for the macroscopic dielectric function:
\begin{equation}
\epsilon(\omega)=\lim_{q \to 0} \frac{1}{[\epsilon^{-1}_{G,G^{\prime}}(q,\omega)]_{{\bf 0},{\bf 0}}}
\end{equation}
which can be rewritten as:
\begin{equation} \label{e2_local_field}
\epsilon(\omega)=\lim_{{\bf q} \to 0}\epsilon_{0,0}({\bf q},\omega)
                -\lim_{{\bf q} \to 0} \sum_{{\bf G},{\bf G}^{\prime}\neq 0} 
                 \epsilon_{0,{\bf G}}({\bf q},\omega)
                 \epsilon^{-1}_{{\bf G},{\bf G}^{\prime}}({\bf q},\omega) 
                 \epsilon_{{\bf G}^{\prime},0}({\bf q},\omega)
\end{equation}
The first term of this equation is the interband contribution to the macroscopic dielectric
function and the second term represents the local field contribution to $\epsilon$.
The determination of  the macroscopic dielectric constant amounts
to the  computation of the  inverse of  
$\epsilon_{{\bf G},{\bf G}^{\prime}}({\bf q},\omega)$. 
Adler and Wiser\cite{Adler} have 
derived, essentially by an extension of the random-phase approximation (RPA), 
an approximation to $\epsilon_{{\bf G},{\bf G}^{\prime}}$ for longitudinal 
fields,
\begin{equation}
\epsilon_{{\bf G},{\bf G}^{\prime}}({\bf q},\omega)=\delta_{{\bf G},{\bf G}^{\prime}}
        - \frac{8 \pi}{\Omega |{\bf q}+{\bf G}| |{\bf q}+{\bf G}^{\prime}|}
        \sum_{{\bf k},n,m}\frac
        {\left [f_{n, {\bf k}-{\bf q}} - f_{m,{\bf k}}\right]
        M^{nm}_{{\bf G}}({\bf k},{\bf q})
        \left [M^{nm}_{{\bf G}^{\prime}}({\bf k},{\bf q})\right]^{*}}
         {E_{n}( {\bf k}-{\bf q})-E_{m}({\bf k}) +\omega + i\delta}
\end{equation}
where $n$ and $m$ are the band indices, $f_{n,{\bf k}}$ is the zero temperature
Fermi distribution, $\Omega$ is the crystal volume and 
$M^{nm}_{{\bf G}}({\bf k},{\bf q})$ are the matrix elements 
\begin{equation} \label{matrix_element}
M^{nm}_{{\bf G}}({\bf k},{\bf q})=
\langle\Psi_{{\bf k}-{\bf q}n}|e^{-i({\bf q}+{\bf G}).{\bf r}}|\Psi_{{\bf k}m}
\rangle
\end{equation}
which are calculated as described in our preceding paper I in the 
context of the GW approximation. In this expression,
the time dependence of the field was assumed to be $e^{-i\omega t}$ and the small
positively defined constant $\delta$ guarantees that the matrix elements of
$\epsilon(\omega)$ are analytic functions in the
half upper plane. Such a
matrix could be separated into an hermitian part $\epsilon^{(1)}_{{\bf G},{\bf G}^{\prime}}$
and an anti-hermitian part $i\epsilon^{(2)}_{{\bf G},{\bf G}^{\prime}}$ according to
\begin{equation}
\epsilon_{{\bf G},{\bf G}^{\prime}}({\bf q},\omega)=
\epsilon^{(1)}_{{\bf G},{\bf G}^{\prime}}({\bf q},\omega)+
i\epsilon^{(2)}_{{\bf G},{\bf G}^{\prime}}({\bf q},\omega)
\end{equation}
with $\epsilon^{(2)}$ for positive $\omega$ given by
\begin{equation}\label{imaginary_part}
\epsilon^{(2)}_{{\bf G},{\bf G}^{\prime}}({\bf q},\omega)=
\sum_{{\bf k},v,c}
\frac{8 \pi^{2}}{\Omega |{\bf q}+{\bf G}| |{\bf q}+{\bf G}^{\prime}|}
M^{vc}_{{\bf G}}({\bf k},{\bf q})
\left [M^{vc}_{{\bf G}^{\prime}}({\bf k},{\bf q})\right]^{*}
\delta\left( \omega -\left[E_{c}({\bf k})-E_{v}( {\bf k}-{\bf q})\right]\right)
\end{equation}
and $\epsilon^{(1)}$ defined by a Kramers Kronig (KK) transform as
\begin{equation} \label{KK}
\epsilon^{(1)}_{{\bf G},{\bf G}^{\prime}}({\bf q},\omega)=
\delta_{{\bf G},{\bf G}^{\prime}}+\frac{2}{\pi}
P\int_{0}^{\infty} d\omega^{\prime}
\frac{\omega^{\prime}\epsilon^{(2)}_{{\bf G},{\bf G}^{\prime}}({\bf q},\omega^{\prime})}
{{\omega^{\prime}}^{2}-\omega^{2}}
\end{equation}
It should be noted here that the matrix elements of $\epsilon^{(2)}$ and $\epsilon^{(1)}$
could be chosen to be real if the inversion is contained in the point group
of the crystal. The calculation of the head element
$\lim_{{\bf q} \to 0}\epsilon^{(2)}_{{\bf 0},{\bf 0}}({\bf q},\omega)$ and of
the wing elements $\lim_{{\bf q} \to 0}\epsilon^{(2)}_{{\bf 0},{\bf G}}({\bf q},\omega)$
necessitate special care if we want to determine the optical properties of semiconductors
when the GW approximation or the scissors-shift approximation is used to 
determine the electronic structure. Instead of handling numerically 
$\lim_{{\bf q} \to 0} M^{nm}_{{\bf 0}}({\bf k},{\bf q})/q$ 
where the quasiparticle 
wave functions $\psi_{{\bf k}n}$ and the quasiparticle energies $E_n({\bf k})$ 
are to be used, it is reasonable to approximate the quasiparticle
wave function with the LDA wave function, and take the limit
analytically\cite{Girlanda}:  
\begin{equation}\label{momentum}
\lim_{{\bf q} \to 0} M^{nm}_{{\bf 0}}({\bf k},{\bf q})/q = 
\widehat{{\bf q}}. \langle n {\bf k} |{\bf p}|m {\bf k}\rangle/
(\epsilon_m({\bf k})-\epsilon_n({\bf k}))
\end{equation}
\noindent
Here  $|n {\bf k}\rangle$ and  $\epsilon_n({\bf k})$ are the LDA wave
functions and energies for band $n$ and wave vector $\bf k$, respectively.  
Indeed, it was shown by inspection that for Si the LDA and the GW wave 
functions have more than 99\% overlap\cite{Hybertsen1}.
If the scissors-shift approximation is used, it is easy to show that
\begin{equation}
\epsilon_{{\bf G},{\bf G^{\prime}}}^{(2)  GW}(\omega)=
\epsilon_{{\bf G},{\bf G^{\prime}}}^{(2)  LDA}(\omega-\Delta)
\end{equation}
where $\Delta$ defines the rigid shift of the conduction bands 
with respect to the valence bands. 

In the above formalism we have neglected the exchange-correlation 
contribution to the dielectric function. The calculation of this 
contribution amounts  basically  to the
determination of the exchange-correlation kernel 
$K_{xc} ({\bf r}, {\bf r^\prime}) = 
\partial^2 E_{xc} / \partial \rho({\bf r}) 
\partial \rho({\bf r^\prime})= {\rm d}V_{xc} /
{\rm d}\rho|_{\rho({\bf r})}  \delta({\bf r} - {\bf r^\prime})$, 
where $E_{xc}$ and $V_{xc}$ are
the exchange-correlation energy and potential,respectively, and  
$\rho({\bf r})$ is the charge
density at $\bf r$. The calculation of $K_{xc}$ is much more complicated 
in all-electron 
than in a pseudo-potential method, since one has to determine the matrix of the Kernel in
the Fourier space. Indeed, the FFT of the exchange-correlation kernel converges very 
slowly  with the number of $\bf G$-vectors in reciprocal space because of the 
oscillating nature of the charge density in real space.
Fortunately, it has been shown that this
kernel contribution  to the total dielectric function is 
small.\cite{Hybertsen2,gavrilenko}

\subsubsection{numerical details}
The size of the dielectric matrix is critical for the convergence of the optical
spectrum. We have found that a size of  $65\times 65$ for all systems studied here
is good for the convergence of the optical spectra, except for LiCl where
the convergence was achieved for a matrix size of $181\times 181$.
The imaginary part of each matrix element 
$\epsilon^{(2)}_{{\bf G},{\bf G}^{\prime}}({\bf q}\to 0,\omega)$ is evaluated 
in energy intervals of 0.1 eV up to 200 eV.
Then the 'real part' $\epsilon^{(1)}$ is deduced via a KK transformation 
defined previously (see Eq. (\ref{KK})).
The linear tetrahedron method\cite{Lehmann,Macdonald} is employed to perform 
the summation over the Brillouin zone which appears in Eq. (\ref{imaginary_part}).
We use 8000 ${\bf k}$-points in the full Brillouin zone to calculate the head element and
1000 ${\bf k}$-points to calculate the wing elements and the body elements. The hermiticity
of $\epsilon^{(2)}_{{\bf G},{\bf G}^{\prime}}({\bf q}\to 0,\omega)$, the time reversal
symmetry and the symmetry properties are used to reduce the number of matrix elements
to be computed. 
\section{Results and discussion }
\subsection{Calculated optical spectra with and without local-field effects}
\begin{figure}
\begin{minipage}{6.0in}
\epsfysize=8.0in
\centerline{\epsfbox{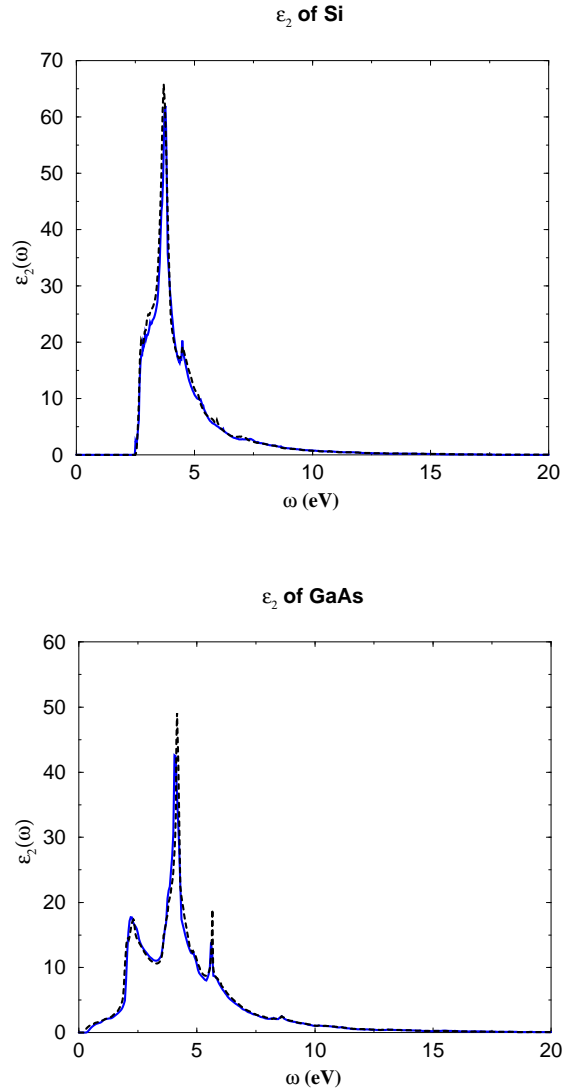}}
\end{minipage}
\caption{
\label{e2_sigaas_fplmto}
Calculated LDA imaginary part of the Dielectric function 
of Si and GaAs versus photon energy using 
PAW method (solid line)
and the FPLMTO method\protect{\cite{aw}} (dashed line). The agreement 
between the two calculations is excellent. This  sets for the first time
the LDA results of the dielectric functions of Si and GaAs.
}
\end{figure}

The dynamical dielectric function of all the semiconductors studied 
here are calculated using Eq. (\ref{e2_local_field}). 
The local-field effects are represented by the second term on the right of this
equation. To test the accuracy of the all-electron PAW method we have computed
the imaginary-part of the dielectric function of Si and GaAs without 
local field and compared them to the full-potential linear muffin-tin (FPLMTO) 
results\cite{aw}. Fig. \ref{e2_sigaas_fplmto} shows that the agreement with the FPLMTO
spectra is excellent. These results are interesting 
because they  set, for the first time, the standard for an  accurate LDA  
dielectric function of Si and GaAs computed  by two different 
all-electron methods. 
This is encouraging since the LMTO method is a state-of-the-art
first principles method for electronic structure, and in comparison, the PAW 
formalism is much simpler, but nevertheless the method doesn't 
loss any accuracy. 

Before presenting  our calculated optical properties of semiconductors, 
we would 
like to mention the different ways we have obtained the optical spectra: 
(1)  We have used our LDA calculated band structure 
to directly compute the optical  spectra.
(2) We have used the so called scissors-operator 
energy shift to the LDA eigenvalues. The value of the shift  
corresponds to our GWA correction of the LDA direct band gap at the
$\Gamma$ point. (3) Finally,  we used  the GWA 
calculated quasiparticle energies across the Brillouin zone. 
All these three types of calculations were produced  with and without 
local-field effects.  

\begin{figure}
\begin{minipage}{6.0in}
\epsfysize=8.0in
\centerline{\epsfbox{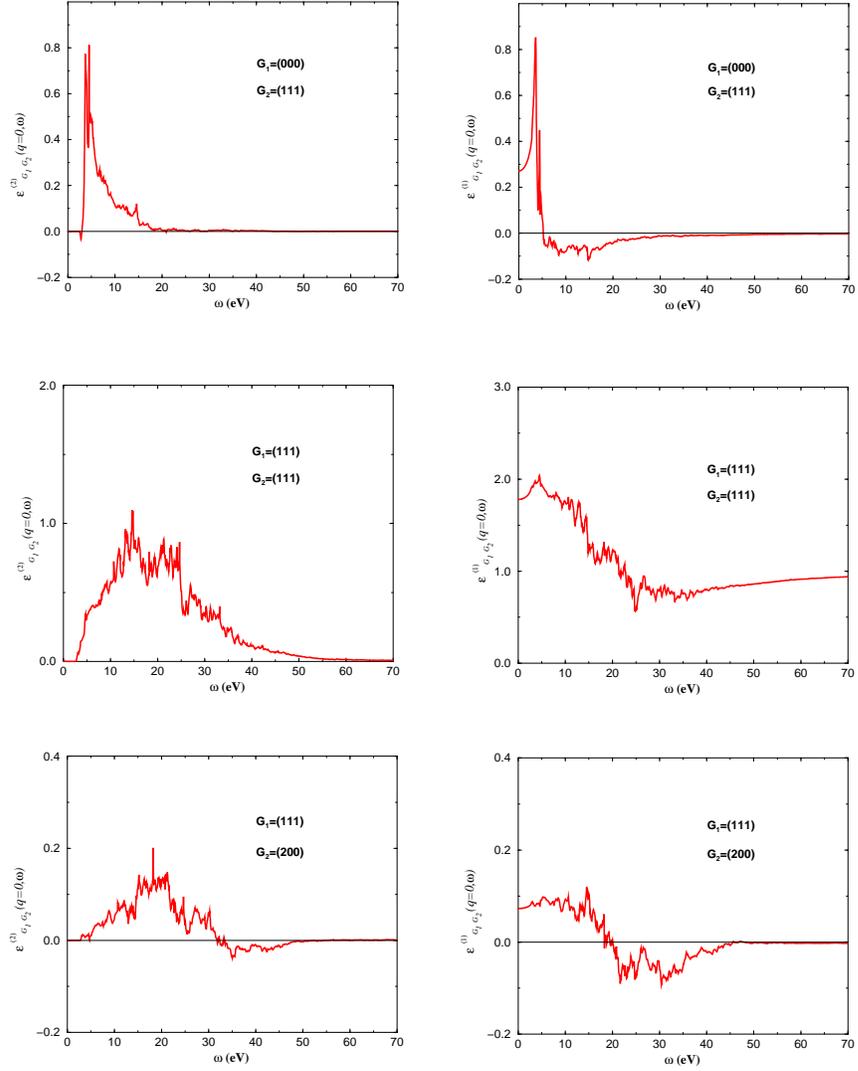}}
\end{minipage}
\caption{
\label{body_wing}
Calculated elements of the real (right column) and imaginary part  (left column)
of the 
symmetrized microscopic dielectric matrix 
$\epsilon({\bf q},\omega)_{{\bf G}, {\bf G^\prime}}$
 of Silicon for the limit $\bf q \rightarrow 0$ and for 
$({\bf G}, {\bf G}^\prime) =$ (000,111), (111,111), and (111,200).   
}
\end{figure}

The accuracy of the macroscopic function depends on the convergence of 
all the elements of the microscopic dielectric matrix.
As stated in the previous section we have found that a matrix of 65 by 65 
$\bf G$-vectors and the use of 200 bands in the interband transitions
produce a well converged $\epsilon(\omega)$, except for LiCl where a
matrix
of $181\times 181$ ${\bf G}$-vectors is used. 
Fig. \ref{body_wing} shows different elements of the microscopic 
dielectric function of Silicon
versus photon energy up to 70 eV. The highest  intensity of these 
elements is at least one order of magnitude smaller than the 
$\epsilon_{(000),(000)}$ element. We compared 
our results to these of Gavrilenko and Bechstedt\cite{gavrilenko} and found that the 
agreement with their results is only at the semi-quantitative level. We 
 are surprised to find that Gavrilenko and Bechstedt have a relatively large 
intensity in the band gap of their  
$Im\epsilon_{(000),(111)}$ and  $Im\epsilon_{(111),(200)}$.

\begin{figure}
\begin{minipage}{6.0in}
\epsfysize=7.0in
\centerline{\epsfbox{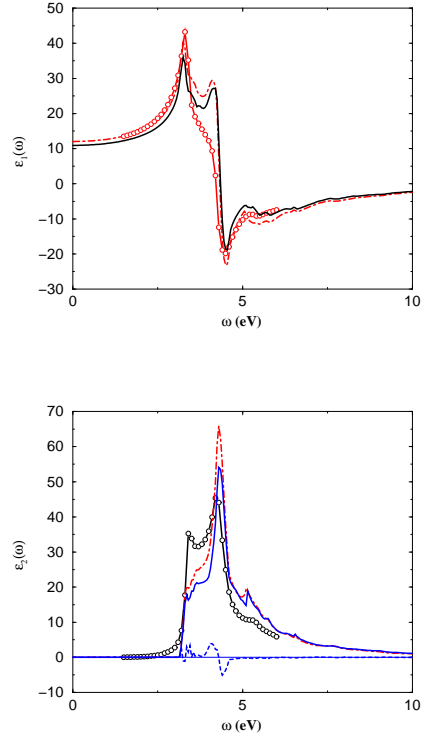}}
\end{minipage}
\caption{
\label{e12_si_LF}
Calculated real and imaginary part of the dielectric function of Si versus
 photon energy.
The  dot-dashed line is our calculation using  a rigid energy shift of 
0.6 eV of the LDA conduction bands, corresponding to our calculated  GW 
band gap correction at the $\Gamma$ point. 
The solid line is  the calculation with a rigid energy shift of the 
conduction bands and including  the local-field effects.
The solid line with open circles is the  experimental 
data\protect{\cite{Aspnes}}. 
The dashed curve is the difference between the calculation
using the quasiparticle energy across the Brillouin zone and 
that  using  the  rigid energy shift. This small difference
justifies the use of the scissors-energy shift for the calculation of the
optical properties.}
\end{figure}

Fig. \ref{e12_si_LF} shows the calculated real and imaginary part of the 
dielectric function of Si versus photon energy up to 10 eV with and 
without LF and QP energy shift. These calculations are
compared to the experimental results of Aspnes and Studna.\cite{Aspnes} 
The dashed curve represent the difference between the calculated
optical spectra using the calculated GW quasiparticle energies and the one
obtained using the scissors-operator energy shift to the LDA eigenvalues.
This spectral difference is small justifying the use of the
scissors-operator for the calculation of the optical spectra of small and
medium gap semiconductors. Because of this small change of the dielectric
function due to the use of the quasiparticle energies, and because of the
high CPU cost in obtaining the quasiparticle energies across the whole
Brillouin zone, all the other small
and medium gap semiconductors are calculated using only the
scissors-operator shift.
Notice that the agreement concerning the peak positions  is 
fortuitous, because the QP energy shift underestimates  
the direct band gap at $\Gamma$ (for details, see paper I). 
A calculation with an energy shift which 
reproduces the experimental band gap will slightly overestimates the peak 
positions by about 0.3 eV, and this overestimation  is valid for all 
semiconductors studied 
here. It is also worth mentionning that our preliminary calculation of the 
excitonic effects shows that in the case of Si the  $E_1$, and $E_2$ peaks are
shifted by about 0.2 eV towards lower photon energies, and the agreement
with experiment concerning the postions and intensities of these peaks is
recovered. 

On the other hand,  in agreement with the empirical 
pseudopotential  (EPP) calculation of Louie, Chelikowsky
and Cohen we have found that the LF effects do not change the position of the
structures present on the optical spectrum\cite{lcc}. 
But the intensities 
of the so called $E_1$ and $E_2$ peaks are reduced, again in agreement with 
the EPP. Thus the local-field effects seem to improve the agreement with 
experiment regarding the intensity of the $E_2$ peak and the structures in 
the higher energy part, and worsen the
agreement with experiment regarding the low energy part 
where the $E_1$ peak is located.  
It is surprising  that our  calculations do not agree well with 
the ab-initio PP calculation of
Gavrilenko and Bechstedt\cite{gavrilenko} which is supposed to be 
similar to the EPM calculation.
The latter calculation  found that while the LF underestimates 
the $E_1$ peak intensity in agreement with our calculation and EPP, 
it overestimates the 
$E_2$ peak intensity in disagreement with our calculation and with EPP.  
The calculated spectrum of Albrecht {\it et al.} \cite{albrecht}, obtained by
solving the Bethe-Salpeter equation, using the PP method, in a 
special limit where only 
local-field effects are included, agrees only 
qualitatively with
our calculation and the EPP results\cite{lcc}. 
Their $E^\prime_1$ structure, which
is located in energy above the $E_2$ structure, has a large intensity and 
disagrees with 
our calculation and  other ab-initio or EPP calculations\cite{aw,lcc}. 
It is then not clear what makes the extra reduction of their $E^\prime_1$
peak  when the excitonic effects are included. 

\begin{figure}
\begin{minipage}{6.0in}
\epsfysize=7.0in
\centerline{\epsfbox{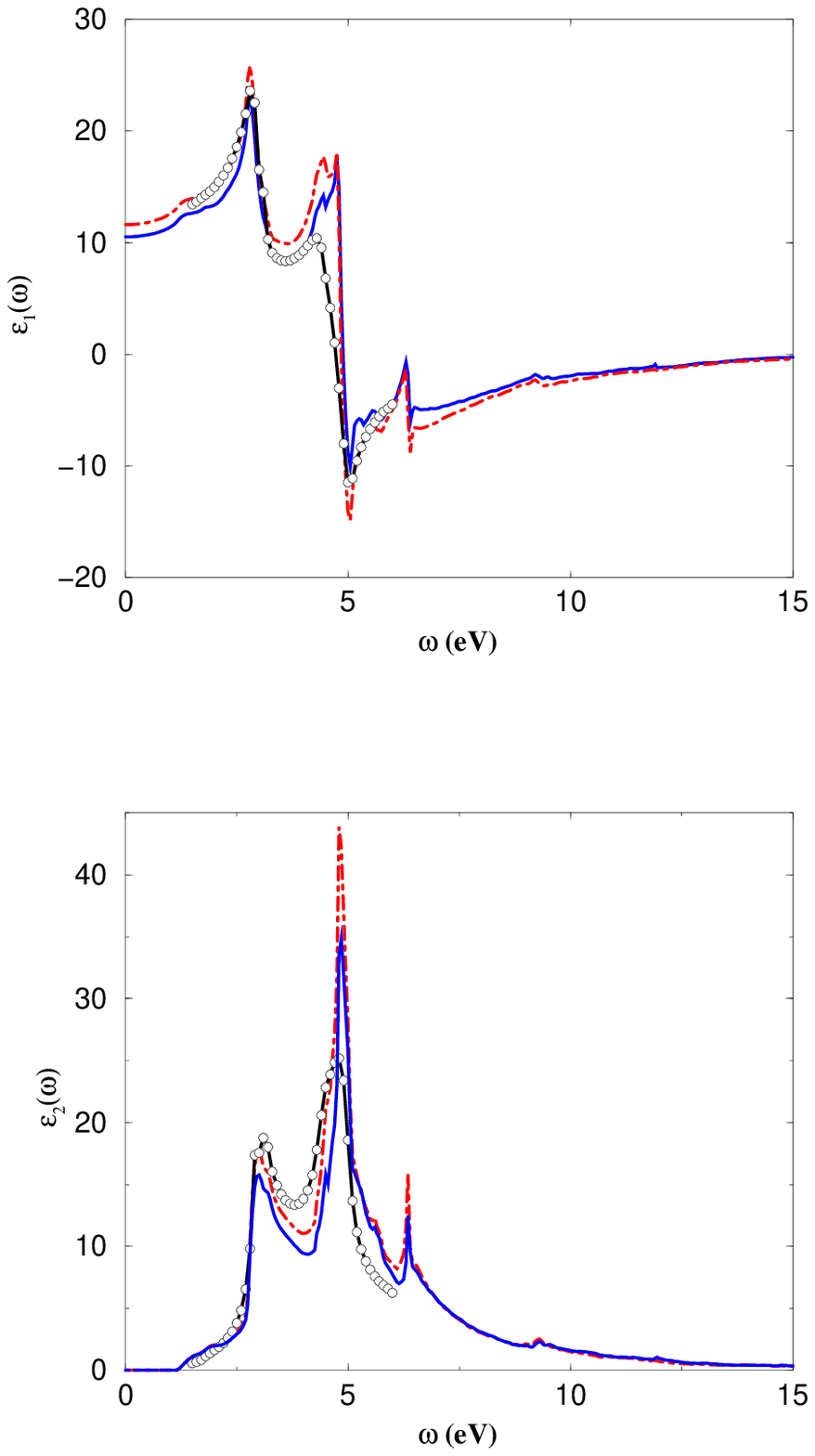}}
\end{minipage}
\caption{
\label{e12_gaas_LF}
Calculated real and  imaginary part of the Dielectric function of GaAs versus
 photon energy.
The dot-dashed line is our  calculation   using  a rigid energy shift of 
0.75 eV of the LDA conduction bands,  corresponding to our calculated  
GW band-gap correction at the $\Gamma$ point.
The solid line is our calculation using  the  rigid energy shift and
including the  local-field effects.
The solid line with open circles is the  experimental spectrum\cite{Aspnes}.
}
\end{figure}

\begin{figure}
\begin{minipage}{6.0in}
\epsfysize=7.0in
\centerline{\epsfbox{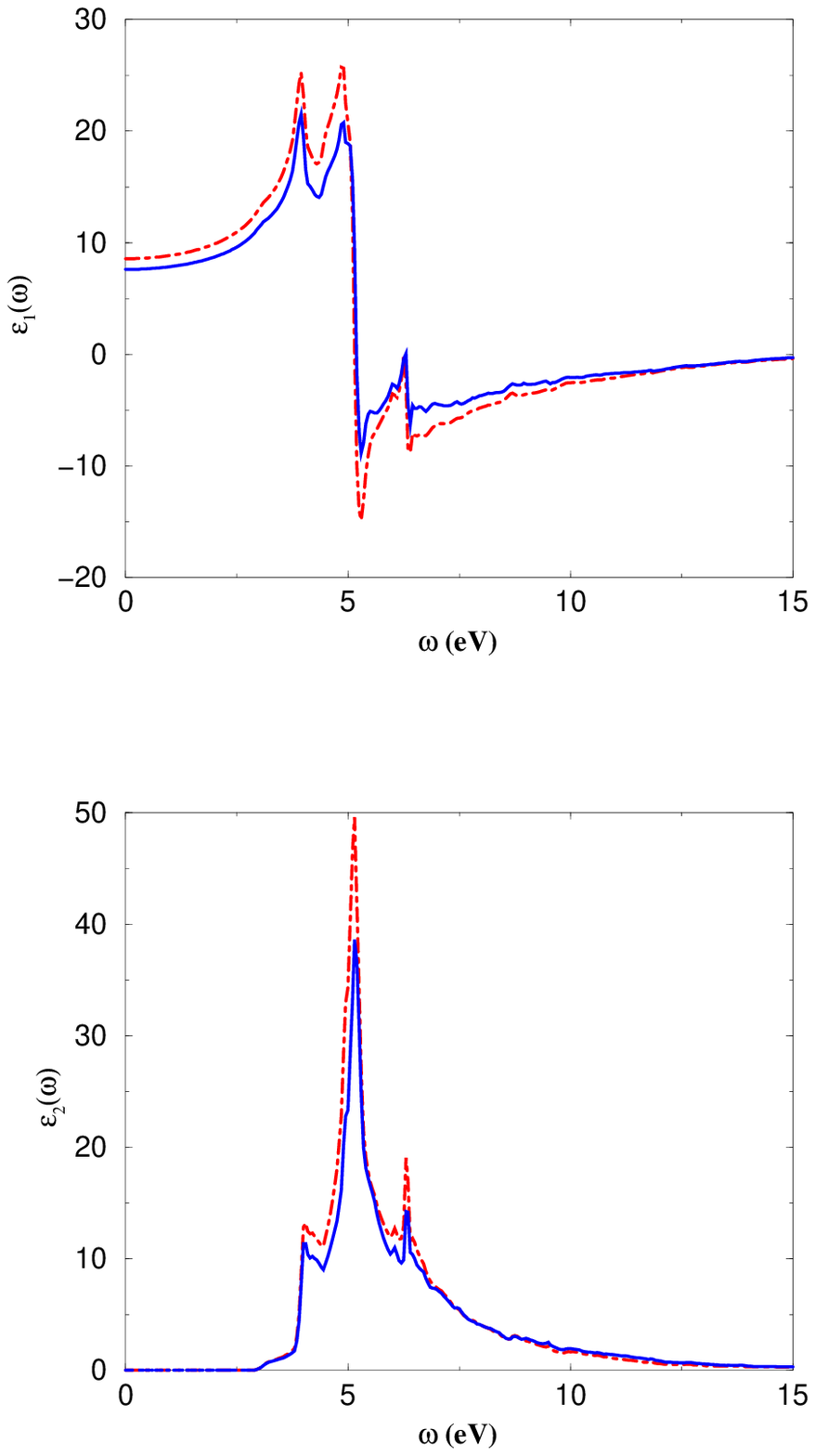}}
\end{minipage}
\caption{
\label{e12_alas_LF}
Calculated real and  imaginary part of the dielectric function of AlAs 
versus  photon energy.
The dot-dashed line is our calculation using  a rigid energy shift of 0.95
eV of the LDA conduction bands, corresponding to our calculated  GW band gap 
correction at the $\Gamma$ point.
The solid line is our  calculation with a rigid energy shift and
 including the  local-field effects.
}
\end{figure}

\begin{figure}
\begin{minipage}{6.0in}
\epsfysize=7.0in
\centerline{\epsfbox{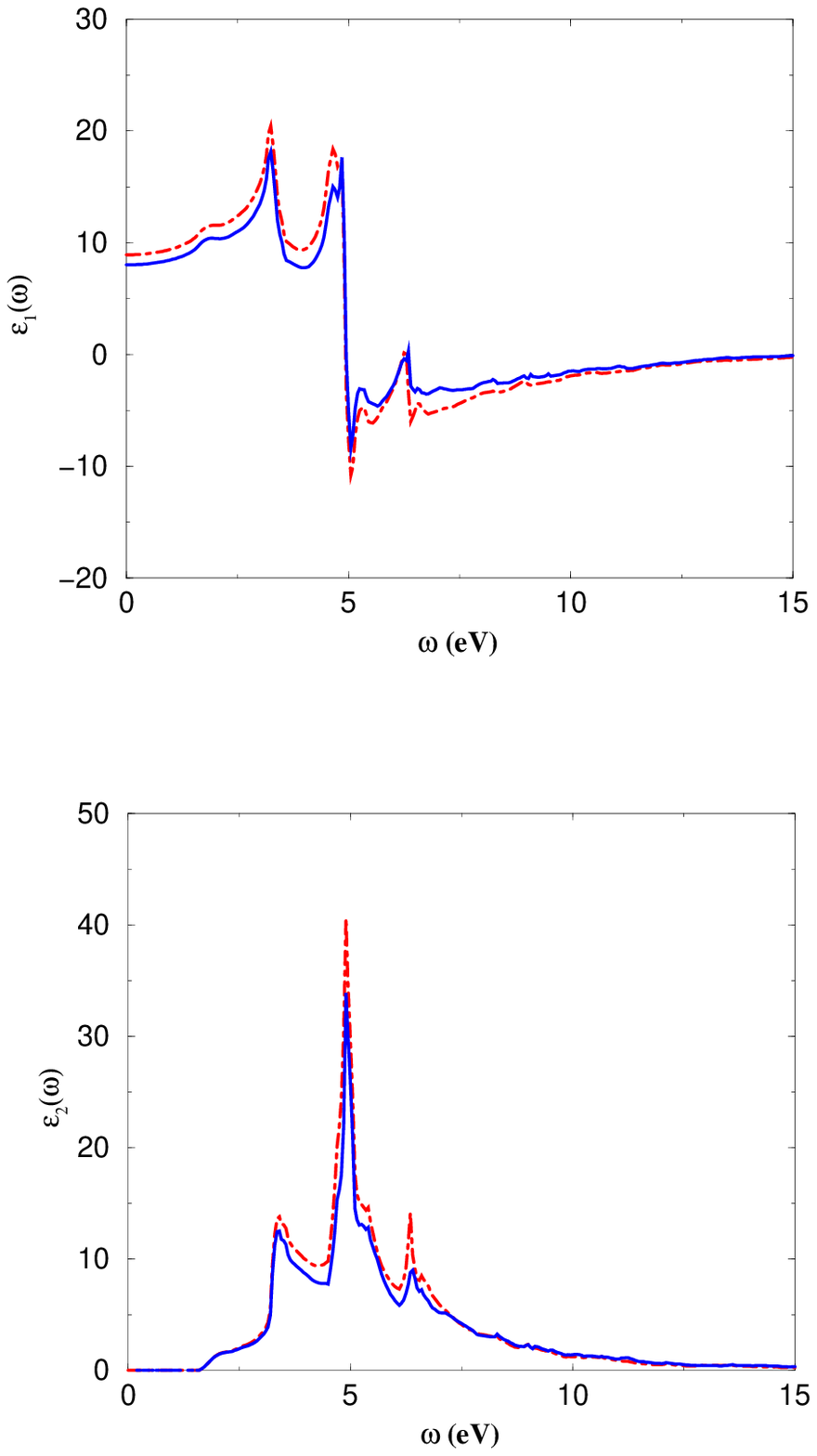}}
\end{minipage}
\caption{
\label{e12_inp_LF}
Calculated real and  imaginary part of the dielectric function of InP 
versus  photon energy.
The dot-dashed line is our calculation using  a rigid energy shift of 0.8
eV of the LDA conduction bands, corresponding to our calculated  GW band gap 
correction at the $\Gamma$ point.
The solid line is our  calculation with a rigid energy shift and
 including the  local-field effects.
}  
\end{figure}

\begin{figure}
\begin{minipage}{6.0in}
\epsfysize=7.0in
\centerline{\epsfbox{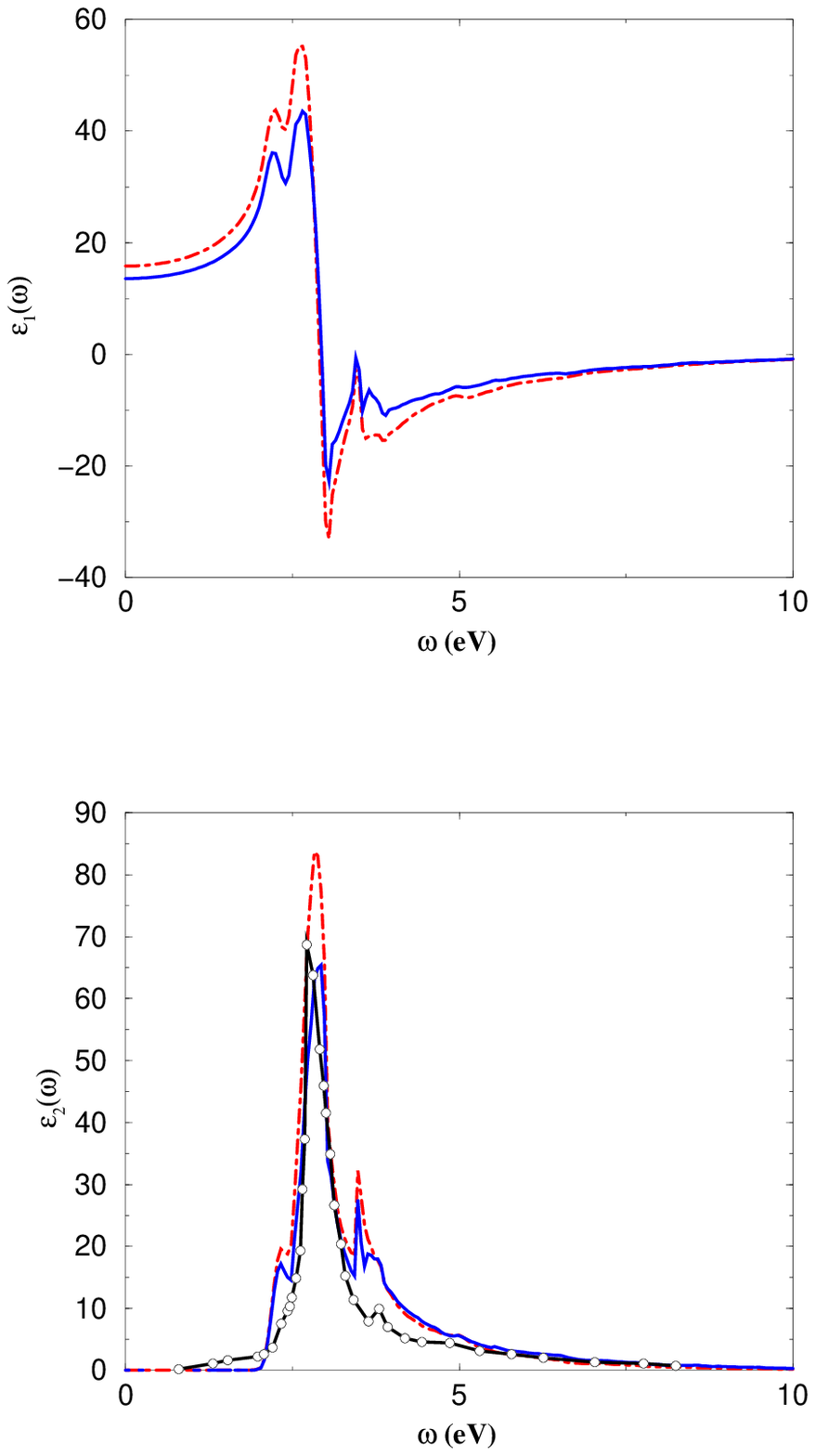}}
\end{minipage}
\caption{
\label{e12_mg2si_LF}
Calculated real and  imaginary part of the dielectric function of Mg$_2$Si 
versus  photon energy.
The dot-dashed line is our calculation using  a rigid energy shift of 0.33
eV of the LDA conduction bands, corresponding to our calculated  GW band gap 
correction at the $\Gamma$ point.
The solid line is our  calculation with a rigid energy shift and
 including the  local-field effects.
The solid line with open circles is the  experimental spectrum\cite{Aspnes}.
}
\end{figure}

\begin{figure}
\begin{minipage}{6.0in}
\epsfysize=7.0in
\centerline{\epsfbox{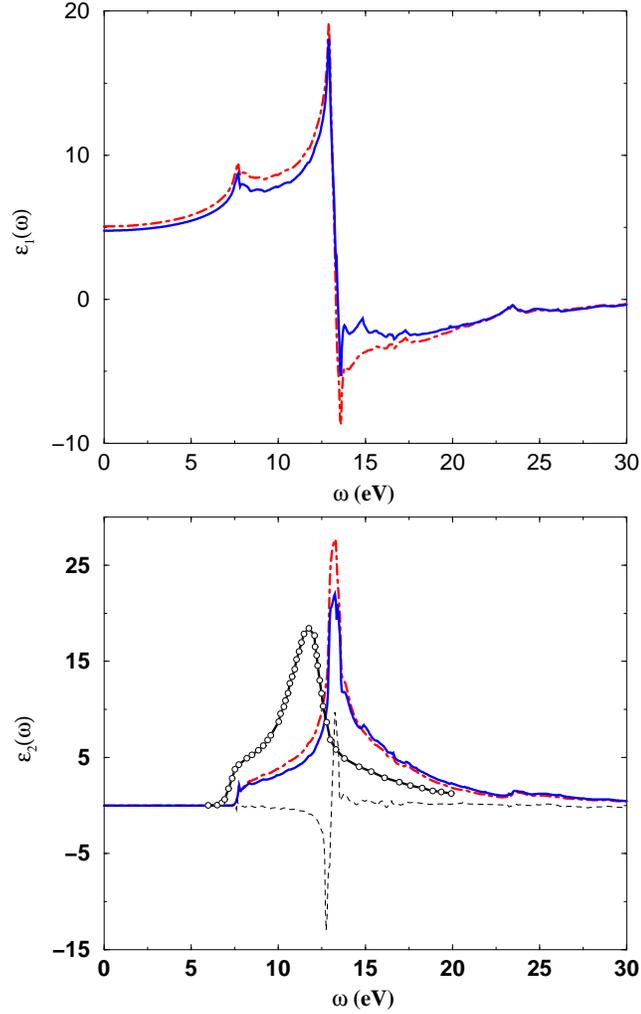}}
\end{minipage}
\caption{
\label{e12_c_LF}
Calculated real and  imaginary part of the Dielectric function of C versus
 photon energy.
The dot-dashed line is our  calculation   using the quasiparticle
energies. 
The solid line is our calculation using  the  quasiparticle energies and
including the  local-field effects.
The solid line with open circles is the  experimental spectrum\cite{papado}.
The  dashed curve represents the difference between the calculations
using the quasiparticle energies and
the LDA calculation with a GW rigid energy shift of 1.9 eV of the
conduction states. This  rigid energy shift corresponds to the GW 
correction of the band gap at the $\Gamma$ point. }
\end{figure}

\begin{figure}
\begin{minipage}{6.0in}
\epsfysize=7.0in
\centerline{\epsfbox{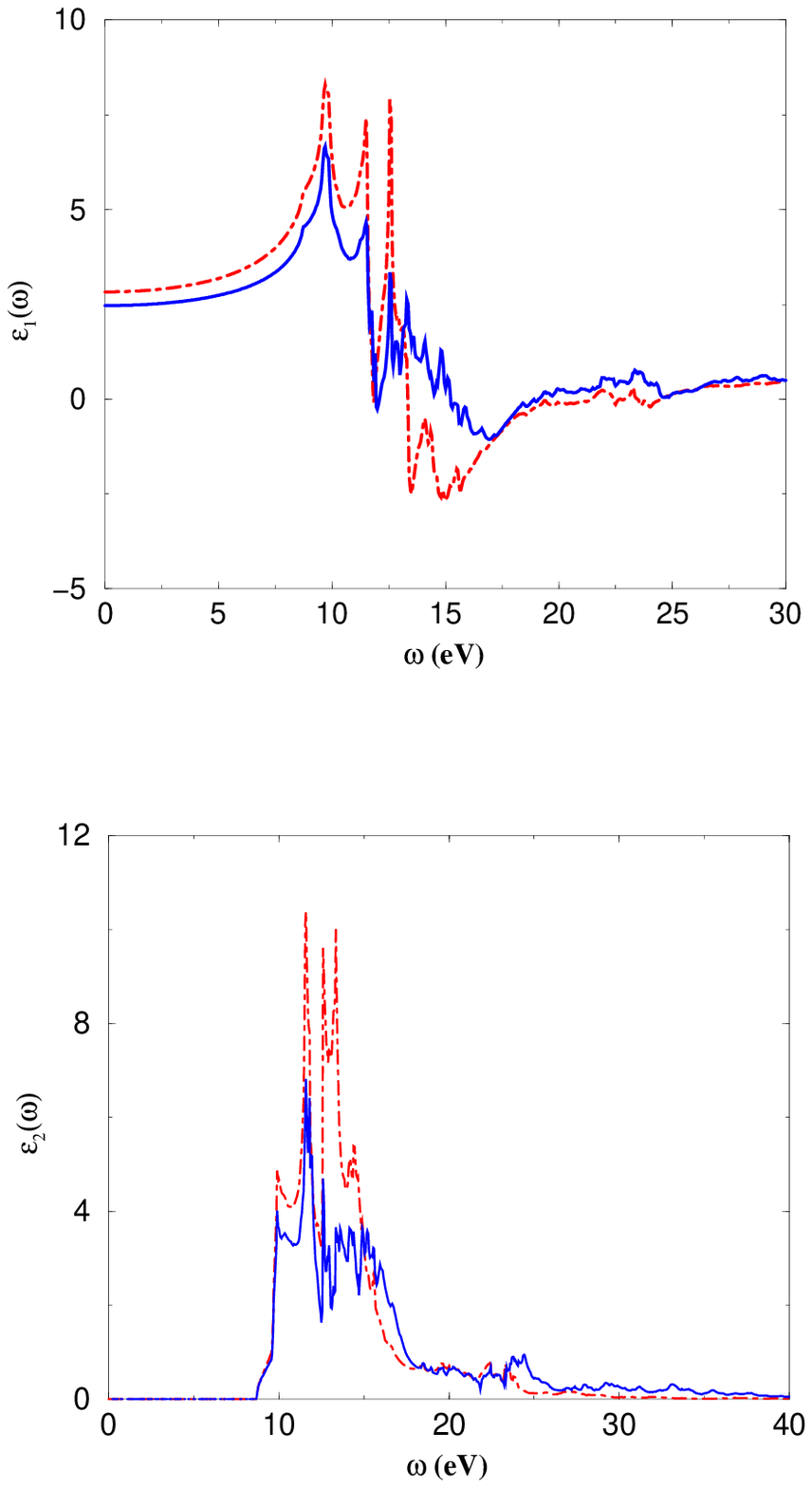}}
\end{minipage}
\caption{
\label{e12_licl_LF}
Calculated real and  imaginary part of the dielectric function of LiCl
versus  photon energy.
The dot-dashed line is our calculation using  a rigid energy shift of 2.8
eV of the LDA conduction bands, corresponding to our calculated  GW band gap 
correction at the $\Gamma$ point.
The solid line is our  calculation with a rigid energy shift and
including the  local-field effects.
}
\end{figure}

We have perfomed similar calculations for GaAs, AlAs, InP, Mg$_2$Si, 
C and LiCl, which we  
present in Figs. \ref{e12_gaas_LF}, \ref{e12_alas_LF}, \ref{e12_inp_LF}, 
\ref{e12_mg2si_LF}, \ref{e12_c_LF}, and  \ref{e12_licl_LF} and 
compare to available experimental results\cite{Aspnes,papado}. 
For all these semiconductor the scissors-operator energy shift, 
corresponding to the 
difference between the quasiparticle and LDA eigenvalues at the
$\Gamma$ point,  is used to produce the
quasiparticle optical spectra, except for C where we have represented
the dielectric function obtained using the GW quasiparticle 
energies at each point in the Brillouin zone. 
In this latter case the difference between the 
optical spectra, calculated using the GW quasiparticle energies and the
scissors-operator energy shift,  is shown by a dashed curve in
Fig. \ref{e12_c_LF}, and it is  found to be  much larger than in the case of 
Si. 
We believe that the same conclusion  should be valid for LiCl, however due to 
the cost of the GW calculation
and the absence of the experimental results we preferred  not to perform the
calculation with the quasiparticle energies. Moreover, the small 
dielectric constants of these two large gap semiconductors indicate that
excitonic effects are important. For example, the large discrepancy between
our calculated spectra of Diamond and experiment can be 
attributed to these effects.

In conclusion, the most important trend of the LF on the optical spectra of 
all types of semiconductors studied here, is that: The $E_1$'s intensity
reduction  disagrees with experiment,  whereas  the $E_2$'s intensity
 reduction  agrees well with experiment whenever available.  
It is interesting to notice that
for the insulator  LiCl the LF effects seem to  reduce substantially the
intensities of the peaks at the low photon  energy. 

\subsection{Electron-energy-loss function}

\begin{figure}
\begin{minipage}{6.0in}
\epsfysize=7.0in
\centerline{\epsfbox{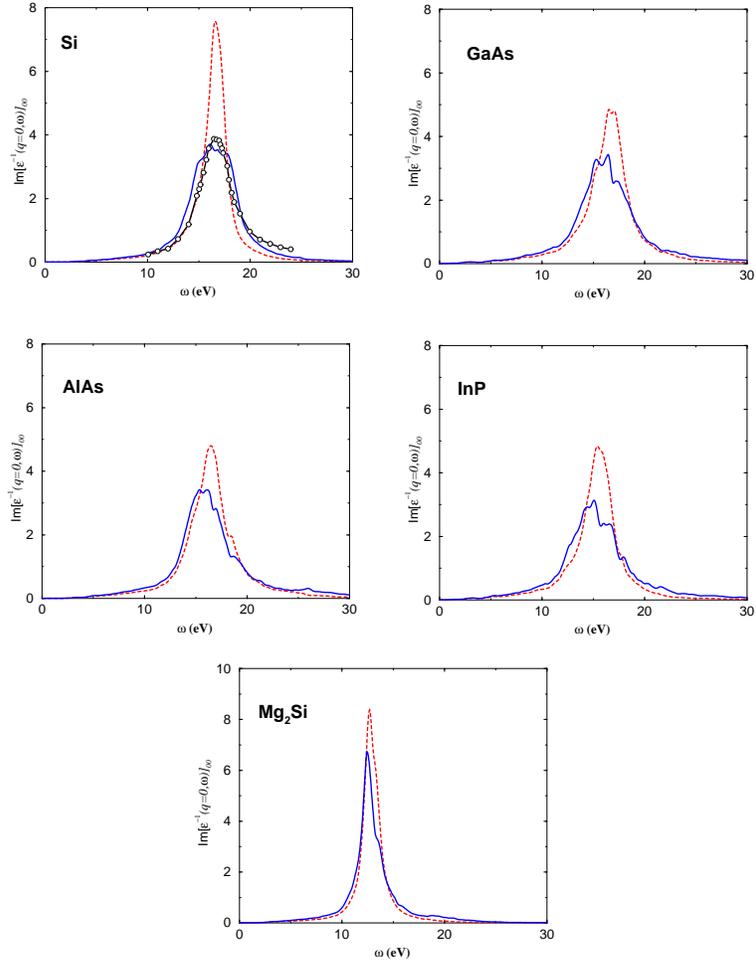}}
\end{minipage}
\caption{
\label{small_gap_loss}
Calculated energy loss function with (solid line) and without local-field 
effects
(dashed line) of small and medium band-gap semiconductors:
Si, GaAs, AlAs, and Mg$_2$Si compared to available experimental 
results\protect{\cite{raether}} (solid line with open circle).  
}
\end{figure}

\begin{figure}
\begin{minipage}{6.0in}
\epsfysize=7.0in
\centerline{\epsfbox{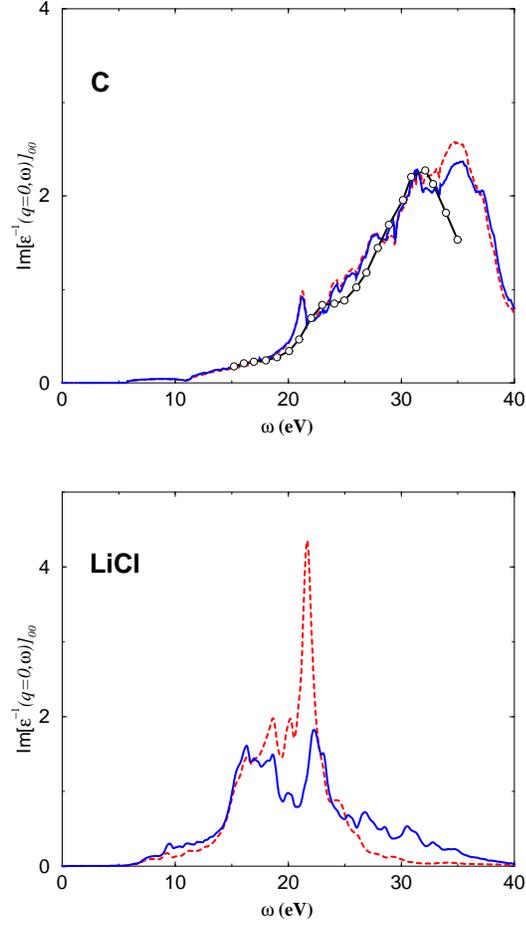}}
\end{minipage}
\caption{
\label{large_gap_loss}
Calculated energy loss function with (solid line) and without local-field 
effects (dashed line) of large band gap semiconductors: C and LiCl
 compared to available experimental results\protect{\cite{phillips}}
(solid line with open circle).  }
\end{figure}

Figs. \ref{small_gap_loss}, \ref{large_gap_loss} show our calculated
electron-energy-loss (EEL) functions $-Im[\epsilon^{-1}(q=0,\omega)]_{0,0}$ 
for small-band-gap semiconductors: Si, GaAs, AlAs,
InP, and Mg$_2$Si and large-band-gap semiconductors: C and LiCl, respectively.
The calculation are done within the LDA with and without the local-field effect.
Whenever possible the calculation is compared to available energy-loss spectra.
The local-field effects seems to improve the agreement with experiment
by reducing significantly the intensity of the main peak. 
The EEL functions of C and specially of LiCl are much complicated. The
LDA C EEL function has two maxima at 31.5 and 34.5 eV and these values 
are shifted to 31.4 and 35.2 eV, respectively, when the LF effects are
included. The experimental curve seems to present only one resonance at 
32 eV. This discrepancy, could be easly due to small  unaccuracy in the
calculated  dielectric function at these high photon energies. 
 
We did not calculate $-Im[\epsilon^{-1}(q=0,\omega)]_{0,0}$ for the 
quasi-particle 
energy because we believe that GWA is not valid at high energies and as 
pointed out in Ref.  \cite{aw} the plasma resonance will be pushed 
towards higher energies in
disagreement with experiment. This because the electronic structure 
at higher energy is
most probably much better described using the  LDA than the GWA because: (1)
at these higher energies the scattering of an electron with the atomic 
potential is small. In this respect, these high  electronic states  
can be obtained from an almost  free-electron theory. 
(2) the plasmon-pole model is not valid at these high
energies. 

Table \ref{plasma_freq} shows the values of the maxima
of the energy-loss function compared to the experimental results obtained 
from the EEL experiments\cite{raether,phillips} and from the measured 
dielectric function\cite{ehrenreich}. The free-electron plasma frequency is 
also shown for comparison.  The plasma resonance 
 of the EEL spectra of Si and GaAs are in good agreement with experiment and 
with the free-electron plasma frequency. The energy electron-energy-loss
function of LiCl is too complicated containing many peaks, and
in the absence of  experimental data we prefered not to show the
values of these  maxima in  table  \ref{plasma_freq}.

\subsection{The static dielectric constant}

\begin{figure}
\begin{minipage}{6.0in}
\epsfysize=7.0in
\centerline{\epsfbox{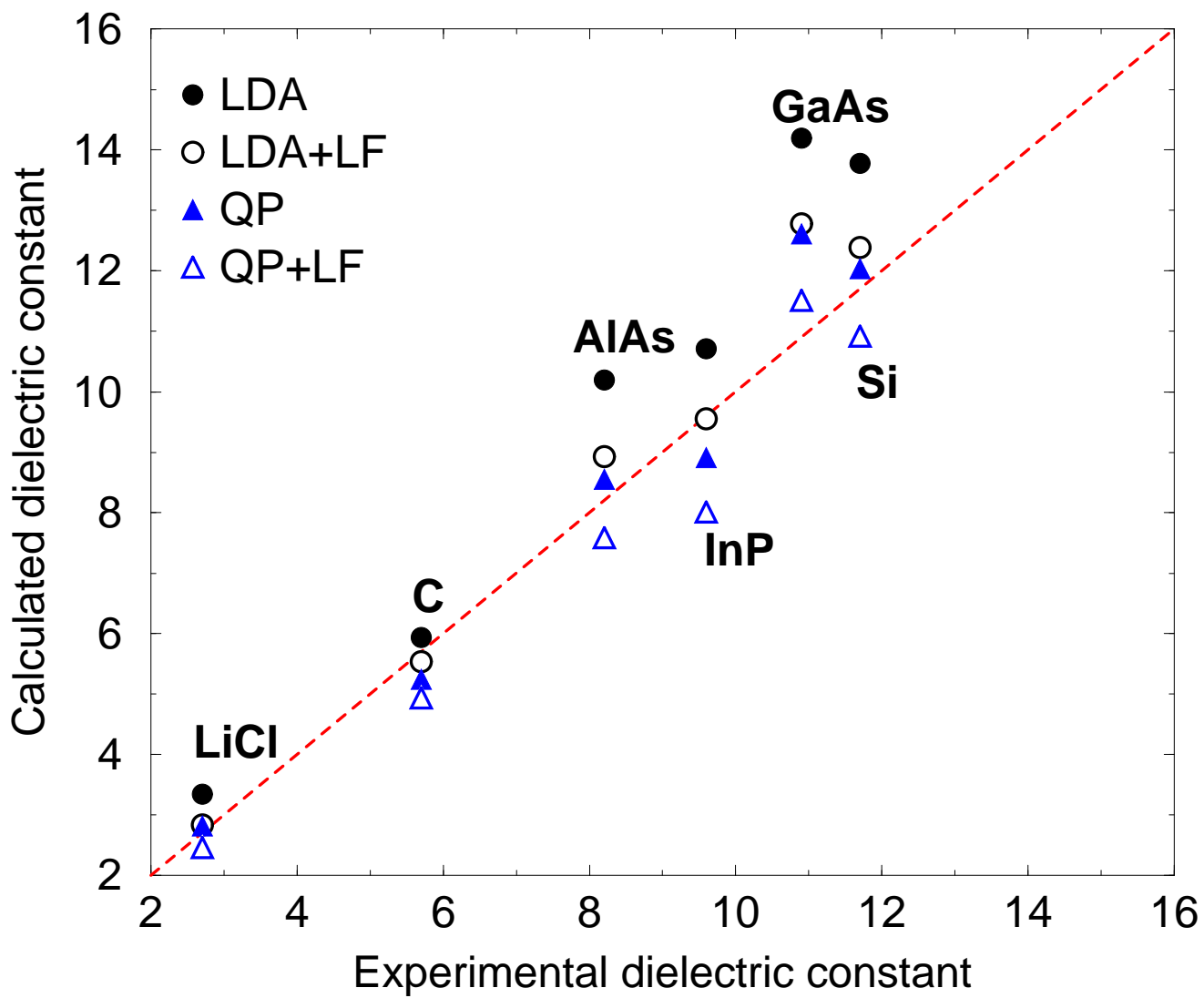}}
\end{minipage}
\caption{
\label{static_df}
Calculated static dielectric function compared to experimental results.
The filled circles represent the LDA values without local-field effects (LF), 
the open circles the LDA values with LF and the up-triangles 
the LDA without LF but with an energy shift corresponding to the GW correction 
of the direct band gap at the $\Gamma$ point,
the empty up-triangles are the LDA values with the GW energy shift and the
LF (see text).
A perfect agreement with experiment is achieved when a calculated value is 
on the dashed line. Notice that when the GW energy shift and the LF are 
included the 
calculation underestimates the static dielectric constant for all these 
semiconductors regardless of the size of the band gap. 
This suggest the importance of the 
excitonic effects which are expected to produce a positive correction 
leading to a better agreement with experiment. 
}
\end{figure}

The static dielectric function $\epsilon_\infty$ with or without local-field 
effects  is computed using the Kramers-Kroenig 
relations.  The calculations were produced using the RPA dielectric function 
and performing analytically  the limit ${\bf q} \rightarrow 0$.
 Table \ref{static_constant}
presents $\epsilon_\infty$ for all semiconductors studied here and compares
them to other calculations\cite{Hybertsen2,Rohlfing,aw,levine1,raynolds,levine2}
and to available experimental results\cite{numerical_data}.  To  
illustrate our data and 
stress the agreement with experiment  we show 
in Fig. \ref{static_df} all our calculated results versus experiment, 
except for
Mg$_2$Si were we are not aware of any  available experimental result.
 A perfect agreement with experiment is achieved when the calculated value
is on the dashed line.  
Because the static dielectric function is a ground
state property, we expect that the calculation with LDA including the 
LF effect should reproduce the experimental results. However, we observe
only an improvement due to this effect. We conclude that the remaining of
the discrepancy with experiment is due to the used of the LDA itself
instead of the full density functional theory. Notice that for large
band gap semiconductors, the LDA results including the LF effects are 
in good agreement with experiment. 
This trend was also observed by other researchers \cite{aw,chen}. 

On the other hand, the 
quasiparticle description of the static dielectric constant, is of
importance, since it will be directly compared to LDA results.
A comparison to experiment will show whether the excitonic effect corrections
are important or not. In this respect we notice that,  when the GWA energy 
shift and LF effects are both included,  the QP
calculation slightly underestimates the static dielectric function for all 
the semiconductors studied here regardless of the size of the band gap or 
the type of semiconductor. 
This suggests, effectively, the importance of the excitonic effects which are 
expected to produce a positive contribution leading to a better agreement 
with experiment.  
This positive contribution arises from  the reduction of 
the optical band gap from its GWA counterpart, and to the transfer of
the force of the oscillator towards lower photon energies. 
It is however interesting to 
remark that since for large band gap materials (C and LiCl) 
the static dielectric function 
within LDA including  LF effects agrees nicely with experiment,
 the excitonic contribution should cancel out the QP correction.  

\section{Conclusion}
We have used our previously implemented GWA within the all-electron 
projected augmented wave method (PAW) to study the optical properties of 
some small, medium and large-band-gap semiconductors: 
GaAs, AlAs, InP, Mg$_2$Si, C and LiCl.
In general, the inclusion of the 
the quasiparticle (QP) energy shift and the local-field (LF) effects 
improves the 
agreement with experiment. In particular, the LF effects reduce the intensities
of the so called $E_1$ and $E_2$ peaks without changing their energy positions.
This reduction of the peak intensity worsen the agreement for the $E_1$ peak but
improves it for the $E_2$ peak.  This trend is  observed for all the 
studied semiconductors and is found to be  in agreement 
 with the empirical pseudopotential  (EPP) calculation of Louie, Chelikowsky
and Cohen \cite{lcc}. The QP energy shift pushes the calculated peaks towards
higher energies in agreement with experiment. However, 
because the calculated GWA energy shift does not always  produce the correct 
experimental band gaps, the agreement of the peak positions with 
experiment in the case of 
Si, GaAs, and Mg$_2$Si is fortuitous.  A calculation using 
an energy shift 
which reproduces the experimental band gap will produce  theoretical peaks
slightly higher in energy compared to experiment.  
On the other hand, our   preliminary calculation of the 
excitonic effects for Si shows that the   $E_1$, and $E_2$ peaks are
shifted by about 0.2 eV towards lower photon energies, and the agreement
with experiment concerning the postions and intensities of these peaks is
is recovered.  Thus, the slight shift of the peaks at higher energy, when
the experimental band gaps are used, is canceled  by  the 
excitonic effects.

The static dielectric function $\epsilon_\infty$ with or without local-field 
effects  is computed using the Kramers-Kroenig 
relations.  The calculation were performed using the RPA dielectric function 
and performing analytically  the limit ${\bf q} \rightarrow 0$.  
Because the static dielectric function is a ground
state property, the calculation using   LDA and including the 
LF effect should reproduce the experimental results. 
In our calculations,  for small and medium band gap semiconductors, we observe
only an improvement due to this effect,  and concluded  that the remaining of
the discrepancy  is due to the use of  the LDA itself
instead of the full density functional theory. However,  for large
band gap semiconductors, the LDA results including the LF effects are 
in good agreement with experiment. 

On the other hand, the QP calculation, when both the  GWA energy shift and 
LF effects are included,   underestimated the static dielectric function for 
all the semiconductors studied here.
This suggests the importance of the excitonic effects which are expected 
to produce a positive contribution.  

\section{Acknowledgment}
We would like to thank P. Bl\"ochl for providing us with his PAW code 
and for useful discussions. Part of this work was done during our
visit to the Ohio State University, and we would like to thank 
J. W. Wilkins and W. Aulbur for useful discussions.  
The Supercomputer time was granted by CINES
on the IBM SP2 supercomputer  (project gem1100). 
%==============================================================================

%                                TABLE 1
%
\begin{table}
\caption{Influence of LF on the energy position of the plasmon peak of the
electron-energy-loss spectra.  Our calculation is compared to available
experimental results and to the free-electron plasma frequency (in eV). 
}
\label{plasma_freq}
\begin{tabular}{lcccc}
Material     &  LDA   & LDA+LF  & Free electron &  Expt.   \\ \hline
Si           &  16.6  & 16.5    & 16.6          &  16.4$^{\rm a}$, 16.9$^{\rm b}$ \\
GaAs         &  16.8  & 16.4    & 15.6          & 14.7$^{\rm a}$         \\
AlAs         &  16.45 & 15.8    & 15.8          &          \\
InP          &  15.5  & 15.0    & 14.8          &           \\
Mg$_{2}$Si   &  12.65 & 12.5    & 13.0          &         \\
C            &  31.5 and 34.5 & 31.4 and 35.2&  31.2   &  32$^{\rm c}$ \\
\end{tabular}
$^{\rm a}$Ref.\protect{\cite{ehrenreich}}, $^{\rm b}$Ref.\protect{\cite{raether}}, 
$^{\rm c}$Ref.\protect{\cite{phillips}}
\end{table}
%%%%%Licl 18.08 eV (free electron)

 %                                TABLE 2
%

\begin{table}
\caption{Influence of LF and QP shifts on the macroscopic dielectric constant
$\epsilon_{\infty}$ compared to other calculations and  to experiment.
\protect{\cite{numerical_data}}}
\label{static_constant}
\begin{tabular}{lccccccc}
Material     &  LDA   &      &LDA+LF  & &QP shift & QP shift+LF & Expt.   \\ \hline
Si           &  13.78 &  13.6$^{\rm a}$,13.8$^{\rm b}$ &12.39   &12.2$^{\rm a}$,12.4$^{\rm b}$ & 12.04    & 10.92       & 11.7    \\
Si           &   &13.75$^f$, 12.8$^{\rm c}$ &   &  &     &        &    \\
GaAs         &  14.23 & 13.1$^c$, 14.17$^d$  &12.78   &  & 11.61    & 10.51       & 10.9    \\
AlAs         &  10.20 &9.5$^{\rm e}$  &8.93    & 8.65$^{\rm e}$  & 8.55     & 7.59        & 8.2     \\
InP          &  10.71 &  &9.55    &  & 8.91     & 8.01        & 9.6     \\
Mg$_{2}$Si   &  17.73 &  &15.22   &  & 15.79    & 13.56       &         \\
C            &  5.94  & 5.5$^{\rm c}$  &5.54    &5.62$^{\rm a}$  & 5.25     & 4.94        & 5.7     \\
LiCl         &  3.35  &  &2.84    & 2.9$^{\rm a}$  & 2.82     & 2.46        & 2.7     \\
\end{tabular}
$^{\rm a}$Ref.\protect{\cite{Hybertsen2}}, $^{\rm b}$Ref.\protect{\cite{levine1}}, 
$^{\rm c}$Ref.\protect{\cite{Rohlfing}}
$^{\rm d}$Ref.\protect{\cite{raynolds}},   $^{\rm e}$Ref.\protect{\cite{levine2}},
$^{\rm f}$Ref.\protect{\cite{aw}}.
\end{table}

%%%%%%%%%%%%%%%%%%%%%%%%%%%%%%%%%%%%%%%%%%%%%%%%%%%%%%%%%%%%%%
\newpage

%%%%%%%%%%%%%%%%%%%%%%%%%%%%%%%%%%%%%%%%%%%%%%%%%%%%%%%%%%%%%
%                                 FIGURE 1
%

%%%%%%%%%%%%%%%%%%%%%%%%%%%%%%%%%%%%%%%%%%%%%%%%%%%%%%%%%%%%%%
\newpage

%%%%%%%%%%%%%%%%%%%%%%%%%%%%%%%%%%%%%%%%%%%%%%%%%%%%%%%%%%%%%
%                                 FIGURE 2
%

%%%%%%%%%%%%%%%%%%%%%%%%%%%%%%%%%%%%%%%%%%%%%%%%%%%%%%%%%%%%%%
\newpage
%%%%%%%%%%%%%%%%%%%%%%%%%%%%%%%%%%%%%%%%%%%%%%%%%%%%%%%%%%%%%
%                                 FIGURE 3
%

%%%%%%%%%%%%%%%%%%%%%%%%%%%%%%%%%%%%%%%%%%%%%%%%%%%%%%%%%%%%%%
\newpage

%%%%%%%%%%%%%%%%%%%%%%%%%%%%%%%%%%%%%%%%%%%%%%%%%%%%%%%%%%%%%
%                                 FIGURE 4
%

%%%%%%%%%%%%%%%%%%%%%%%%%%%%%%%%%%%%%%%%%%%%%%%%%%%%%%%%%%%%%%
\newpage

%%%%%%%%%%%%%%%%%%%%%%%%%%%%%%%%%%%%%%%%%%%%%%%%%%%%%%%%%%%%%
%                                 FIGURE 5
%
%
%
%%%%%%%%%%%%%%%%%%%%%%%%%%%%%%%%%%%%%%%%%%%%%%%%%%%%%%%%%%%%%%
\newpage

%%%%%%%%%%%%%%%%%%%%%%%%%%%%%%%%%%%%%%%%%%%%%%%%%%%%%%%%%%%%%
%                                 FIGURE 6
%
%
%%%%%%%%%%%%%%%%%%%%%%%%%%%%%%%%%%%%%%%%%%%%%%%%%%%%%%%%%%%%%%
\newpage

%
%
%%%%%%%%%%%%%%%%%%%%%%%%%%%%%%%%%%%%%%%%%%%%%%%%%%%%%%%%%%%%%%
\newpage

%%%%%%%%%%%%%%%%%%%%%%%%%%%%%%%%%%%%%%%%%%%%%%%%%%%%%%%%%%%%%
%                                 FIGURE 7
%

%%%%%%%%%%%%%%%%%%%%%%%%%%%%%%%%%%%%%%%%%%%%%%%%%%%%%%%%%%%%%%
\newpage

%%%%%%%%%%%%%%%%%%%%%%%%%%%%%%%%%%%%%%%%%%%%%%%%%%%%%%%%%%%%%
%                                 FIGURE 8
%

%%%%%%%%%%%%%%%%%%%%%%%%%%%%%%%%%%%%%%%%%%%%%%%%%%%%%%%%%%%%%
\newpage
%                                 FIGURE 9
%
%%%%%%%%%%%%%%%%%%%%%%%%%%%%%%%%%%%%%%%%%%%%%%%%%%%%%%%%%%%%%
\newpage
%                                 FIGURE 10
%
%%%%%%%%%%%%%%%%%%%%%%%%%%%%%%%%%%%%%%%%%%%%%%%%%%%%%%%%%%%%%
\newpage
%                                 FIGURE 11
%

\end{document}